
\input phyzzx.tex

\tolerance=1000

\twelvepoint
\normalbaselineskip=16pt


\REF\VA{D.V. Volkov and V.P. Akulov, 
{\it Is the Neutrino a  Goldstone Particle}, 
Phys. Lett. {\bf B46} (1973) 109.}

\REF\GL{Yu, A, Golfand and E.P. Likhtman, {\it Extension of the Algebra
of Poincare Group Generators and Violation of P Invariance},   
JETP Lett. {\bf 13} (1971) 323.}

\REF\SS{A. Salam and J. Strathdee, {\it On Golstone Fermions}, 
Phys. Lett. {\bf B49} (1974) 465.}

\REF\dF{B. de Wit and D.Z Freedman, 
{\it Phenomenology Of Goldstone Neutrinos}, 
Phys. Rev. Lett. {\bf 35} 827, 1975.}

\REF\RSh{V.A. Rubakov and M.E. Shaposhnikov, 
``Do We Live Inside A Domain Wall?'', 
Phys. Lett. {\bf B125} (1983) 136}

\REF\DS{G. Dvali and M. Shifman,
``Dynamical Compactification as a Mechanism of Spontaneous 
Supersymmetry Breaking'',
Nucl. Phys. {\bf B504} (1997) 127, hep-th/9611213.}

\REF\AH{N. Arkani-Hamed, S. Dimopoulos and G. Dvali
``The Hierarchy Problem and New Dimensions at a Millimeter'',
Phys. Lett. {\bf B429} (1998) 263, hep-ph/9803315.}

\REF\RS{L. Randall and R. Sundrum, ``An Alternative to
  Compactification'', 
Phys. Rev. Lett. {\bf 83} (1999) 4690, hep-th/9906064.}

\REF\LW{N.D. Lambert and P.C. West, {\it M-Theory and Hypercharge},
JHEP, {\bf 9909} (1999) 021, hep-th/9908129.}

\REF\W{E. Witten, {\it Solutions of Four-dimensional Field Theories 
Via M Theory}, Nucl. Phys. {\bf B500} (1997) 3, hep-th/9703166.}

\REF\HLW{P.S. Howe, N.D. Lambert and P.C. West, 
{\it Classical M-Fivebrane Dynamics and Quantum N=2 Yang-Mills}, 
Phys. Lett. {\bf B418} (1998) 85, hep-th/9710034.}

\REF\ADD{N. Arkani-Hamed, S. Dimopoulos and G. Dvali, {\it 
Phenomenology, Astrophysics and Cosmology of Theories with Sub-Millimeter
Dimensions and TeV Scale Quantum Gravity}, 
Phys. Rev {\bf D59} (1999) 096004, hep-ph/9807344.}

\REF\GKL{G. Gibbons, R. Kallosh and A. Linde, {\it Brane World Sum
    Rules }, hep-th/0011225}

\REF\GSW{J. Goldstone, A. Salam and S. Weinberg, {\it Broken
    Symmetries},  Phys. Rev. {\bf 127} (1962) 965.}

\REF\J{R. Jackiw, {\it Quantum Meaning of Classical Field Theory}, 
Rev. Mod. Phys. {\bf 49} (1977) 281.}

\REF\S{T.H.R. Skyrme, Proc. Roy. Soc. {\bf A260} (1961) 127.}

\REF\ANW{G.S. Adkins, C.R. Nappi and E. Witten, {\it Static
   Properties of Nuclons in the Skyrme Model}, {\bf B228} (1983) 552.}

\REF\MK{M.P. Mattis and M. Karliner, {\it Baryon Spectrum of the
    Skryme Model}, Phys. Rev. {\bf D31} (1985) 2833.} 

\REF\HLWthree{P.S. Howe, N.D. Lambert and P.C. West, 
{\it The Threebrane Soliton of the M-fivebrane}, 
Phys. Lett. {\bf B419} (1998) 79, hep-th/9710033.}

\REF\GP{G.W. Gibbons and G. Papadopoulos, 
``Calibrations and Intersecting Branes'', Commun. Math. Phys. 
{\bf 202} (1999) 593, hep-th/9803163.}

\REF\GLWone{J.P. Gauntlett, N.D. Lambert and P.C. West, {\it Branes
    and Calibrated Geometries}, Comm. Math. Phys. {\bf 202} (1999)
    517,  hep-th/9803216.}

\REF\QMW{B.S. Acharya, J.M. Figueroa-O'Farrill and B. Spence, 
``Branes at Angles and Calibrated Geometry'',
JHEP {\bf 9804} (1998) 012, hep-th/9803260.}

\REF\AKT{I. Antoniadis, E. Kiritsis and T.N. Tomaras, {\it A D-Brane
Alternative to Unification}, Phys. Lett. {\bf B486} (2000) 186,
hep-ph/0004214.}

\REF\AIQU{ G. Aldazabal,  L.E. Ibanez,  F. Quevedo and A.M. Uranga, 
{\it D-Brane at Singularities: A Bottom Up Approach To The String 
Embedding Of
The Standard Model}, JHEP {\bf 0008} (2000) 002,  hep-th/0005067.}

\REF\West{P.C. West, {\it Automorphisms, Non-Linear Realizations and
    Branes},
hep-th/0001216.}

\REF\HLWone{P.S. Howe, N.D. Lambert and P.C. West, 
{\it The Self-Dual String Soliton}, 
Phys. Lett. {\bf B515} (1998) 203, hep-th/9709014.}

\REF\GLW{J.P. Gaunlett, N.D. Lambert and P.C. West, 
{\it Supersymmetric Fivebrane Solitons}, 
Adv. Theor. Math. Phys. {\bf 3} (1999) 91, hep-th/9711024.}

\REF\LWm{
N.D. Lambert and P.C. West , {\it Monopole Dynamics from the M-Fivebrane} 
Nucl. Phys. {\bf B556} (1999) 177, hep-th/9711025.}

\REF\G{J.P. Gauntlett, {\it Membranes on Fivebranes}, hep-th/9906162.}

\REF\St{A. Strominger, {\it Open p-Branes}, Phys. Lett. {\bf B383}
  (1996) 44,  hep-th/9512059.}

\REF\T{P.K. Townsend, {\it D-Branes from M-Branes}, Phys. Lett. {\bf
    B373} 68, hep-th/9512062}

\REF\HT{C.M. Hull and P.K. Townsend,  {\it Unity of Superstring Dualities}, 
Nucl. Phys. {\bf B438} (1995) 109, hep-th/9410167.}

\REF\HS{G.T. Horowitz and A. Strominger, {\it Black Strings and
    p-Branes}, Nucl. Phys. {\bf B360}, (1991) 197.}

\REF\P{J. Polchinski, {\it D-branes and  Ramond-Ramond Charges}, 
Phys. Rev. Lett. {\bf 75} (1995) 4724, 
hep-th/9510017.}

\REF\Fayet{P. Fayet, {\it ABout R Parity and the Supersymmetric
    Standard Model}, in Shifman, M.A. (ed.): 
``The Many Faces of the Superworld'', hep-ph/9912413.}

\pubnum={KCL-TH-00-65\cr hep-th/0012121}
\date{December 2000}

\titlepage

\title{\bf Goldstone-Soliton Interactions and Brane World Neutrinos}

\vskip 24pt

\centerline{N.D. Lambert 
and P.C. West\foot{lambert,pwest@mth.kcl.ac.uk}}

\address{Department of Mathematics\break
         King's College, London\break
         WC2R 2LS\break
         England\break
         }

\abstract
We discuss the interactions of Goldstone particles with solitonic
states. We observe that, contrary to the familiar 
situation in the vacuum sector, 
the Goldstone particles can have non-derivative
interactions with the solitons. This result is applied to
brane physics and in particular leads to the possibility that 
neutrinos in brane world scenarios are Goldstone particles 
for broken supersymmetry. 
\endpage


\chapter{Introduction}

One of the earliest developments in supersymmetry was the 
paper by  Volkov and Akulov [\VA] which
found a non-linear realisation of the supersymmetry  algebra that was 
reported in the early paper of Golfand and Likhtman [\GL]. 
This quantum field theory 
described a massless Fermion interacting with itself through derivative 
interactions. By analogy with the case for internal symmetries, the 
the Fermion described by this non-linear realisation was to be
thought of as the Goldstone particle that should arise if supersymmetry 
is spontaneously broken. These authors then proposed that the Goldstone 
particle corresponding to the spontaneous breaking of supersymmetry 
was the neutrino. It was later shown [\SS] that supersymmetry did
indeed obey a Goldstone 
theorem and  spontaneously broken supersymmetry does
inevitably lead to a massless 
particle of spin 1/2, the Goldstino. 
However, a problem arose with its interpretation as 
a neutrino. For the case of internal symmetries it was well-known that 
the Goldstone particles had only derivative interactions with themselves 
and all the other particles in the theory. It was shown [\dF] that 
this theorem also applied to the Goldstino but it was known from
experiment that the neutrino had non-vanishing cross  sections in the limit
of zero momentum with the other leptons. 
Thus the hypothesis that the neutrino was a Goldstino was seen to be in 
contradiction with experiment and laid to rest.  

This failure posed a problem for model building using spontaneously broken 
supersymmetry since this mechanism inevitably lead to a Goldstino and 
this must be a massless particle that had not been observed so far. Later,  
mainly motivated by the tight pattern of masses that occur  
 when spontaneously breaking supersymmetry 
in theories of rigid supersymmetry,  model builders turned to theories 
that involved supergravity. In this case,  supersymmetry is locally 
realised, so that the corresponding Goldstino will be absorbed by the 
gravitino and 
the apparent naive  contradiction with  observation
is removed. 

Within the past two years a significant amount of research has concentrated
on the idea that branes in a higher dimensional spacetime are
relevant for phenomenology (for example see [\RSh,\DS,\AH,\RS,\ADD]). 
A main motivation for
the recurrence of this idea is that in string theory Yang-Mills fields
are naturally confined to the worldvolume of branes. In these scenarios
gravity is still a  bulk field in all of spacetime. 
In this way of viewing things, the world we observe may emerge from 
a fundamental theory in a very different way than from the standard
compactification methods. 

A specific model of this type that seems natural from the point of view of
M-theory is to consider an M-fivebrane
which is wrapped on a two-cycle in a compactification of M-theory 
leaving a four dimensional worldvolume that is identified with 
Minkowski space [\LW]. For example M-theory compactified on a manifold of
$G_2$ holonomy has four-dimensional $N=1$ supersymmetry. 
The supersymmetries of M-theory can be  broken by the wrapping or 
self-intersection of the fivebrane and by the background spacetime.  
One appealing possibility is to break all the supersymmetries and to try to 
find a low energy theory which is just the standard model and 
no more [\LW].

In any  brane world scenario it is natural to imagine that some supersymmetry
is preserved by the bulk spacetime but is broken by the brane on which
the standard model lives. By analogy with  the Goldstone theorem one 
would expect to see massless Fermionic fields on the brane corresponding
to the breaking of bulk supersymmetry by the brane. 
As already mentioned, normally in supergravity
the spin 1/2 Goldstinos are eaten by the gravitinos in a super-Higg's
mechanism. Such a mechanism seems implausible in the case of D-branes 
however 
since the spin 1/2 Goldstinos are confined to the brane whereas the 
gravitini propagate in the bulk spacetime. 

Therefore an apparently robust
feature of these  brane world scenarios is the prediction of massless, 
uncharged Goldstone Fermions. In this way one is naturally returned 
to the idea of Volkov and Akulov
but then also into its conflict with experiment.

We note that 
an interesting Higg's mechanism does exist for smooth
domain walls whose  transverse space is a circle [\ADD]. In particular
the lowest mode of the Kaluza-Klein vector field in  
the bulk eats the Goldstone boson corresponding to broken
translational symmetry around the circle.
It is natural to also 
expect that this mechanism holds for the case of supersymmetries, so that the
would-be Goldstone Fermions are eaten by the lowest mode of the bulk 
gravitini.
However the topological stability of such domain walls is questionable
since  continuity and perodicity of the scalars around the
compact direction imply that the domain wall has no 
topological charge (see for example [\GKL]). 
In any case, in this paper we hope to convince the reader that there is an
alternative resolution to the Goldstino puzzle in which the appearance
of  massless, uncharged Fermions such as neutrinos is rather natural.

There is another example in which the purely derivative interactions of 
Goldstone modes needs to be questioned. We recall that an
M-fivebrane wrapped on a non-compact Riemann Surface can be
related by M-theory/Type IIA duality to $N=2$ Yang-Mills gauge theory 
[\W].
In particular the low energy dynamics of M-fivebrane precisely 
reproduce the Seiberg-Witten effective action for $N=2$ Yang-Mills 
theory [\HLW]. 
The massless Fermions which appear in the M-fivebrane
effective action are Goldstinos for broken supersymmetry and
these must be identified with the massless Fermions that appear 
in the $N=2$ Yang-Mills theory
after the gauge group is spontaneously broken. However, in the full
$N=2$ Yang-Mills Lagrangian these massless Fermions have 
non-derivative couplings to the charged states (e.g. the $W^\pm$). 
Therefore  M-theory/Type IIA duality implies that these Goldstinos
have non-derivative interactions with other states on the M-fivebrane.
Indeed, even without assuming such a  duality, it is certainly the
case that the low energy
effective action for the Goldstino modes of the M-fivebrane is
identical to the effective action of a theory in which these Fermions
have non-derivative interactions.

There is an important connection between these two cases. In the
low energy equations of motion of an M-fivebrane (and D-branes too)
only the field strength, and not the gauge field, appears. Thus none of
the low energy states are charged. All charged states that
appear in the low energy dynamics of branes must arise as soliton solutions
[\LW].
Indeed another related question is how do these  charged 
soliton states couple to the gauge potential.
It is now clear that a way out of both of these dilemmas is to postulate
that Goldstone modes can have non-derivative interactions with soliton
states.  
Therefore in this paper we will explicitly exhibit this mechanism for
a general theory. We then
argue that the goldstino modes corresponding to the breaking of
supersymmetry by the brane will indeed have non-derivative couplings
to the charged states that also live on the brane.

The rest of this paper is organised as follows. In the next section 
we will briefly review Goldstone's theorem, quantisation about 
soliton solutions and apply these ideas to the case of branes. 
In section three we will consider the case of theories which
simultaneously
admit both soliton solutions and Goldstone modes, the best known
example perhaps being the Skyrme model. In particular we
will explicitly demonstrate a mechanism for non-vanishing  Goldstone/soliton
scattering at zero momentum (although this does not occur in the
Skyrme model). In section four we will then focus  the
general discussion to the specific case of the M-fivebrane. Here we will
discuss the resolution of the apparent contraction with type IIA
string  theory and in addition discuss the non-derivative 
interactions that worldvolume Goldstinos have with charged states
in  phenomenological brane models. We note here that in this paper
we will give general arguments to establish the existence and origins of
non-derivative interactions
between Goldstone particles and solitons. However, we will not provide
a
detailed analysis for all cases, such as broken supersymmetries.
We expect that analogous result will follow this and other cases
by a straightforward extension of the discussion presented here. 


\chapter{Branes and Goldstone Particles}

In this paper we will mainly work in $D$ dimensions. We use $\mu,
\nu=0,1,2,,...,D-1$. A $D$-vector will be denoted by $k^\mu$ or
just $k$. We will also need to consider the purely spatial components
of a vector which we denote by $\vec {k}$. We use the letters
$i,j,k,...$ to label the various fields that appear and $A,B,C...$ to
label any global symmetries.

Although it is widely known, it will be helpful if we summarise an
elementary proof of Goldstone's theorem in the case of an internal 
symmetry which is spontaneously broken.
Suppose that we consider the effective action $\Gamma[\phi]$. 
Let us further suppose that the effective action is invariant under a 
symmetry $\delta\phi^i = \omega^A f_A^i(\phi)$ so that
$$
\int d^Dx {\delta \Gamma \over \delta \phi^i}\omega^A f_A^i(\phi) =
0 \ .
\eqn\eai
$$
If consider the case of zero momentum on all external legs 
then $\Gamma[\phi]=-V(\phi)$ is just the
effective potential and is independent of $x^\mu$. Therefore
we can drop the integral from \eai. Differentiating 
with respect to $\phi^j$ we find, at zero 
momentum, 
$$
{\partial^2 V\over \partial \phi^i\partial\phi^j}\omega^A f_A^i(\phi) 
+{\partial V\over \partial \phi^i}\omega^A
{\partial f_A^i\over \partial\phi^j} =0 \ . 
\eqn\variation
$$
Upon setting $\phi^i=\phi^i_0$, where $\phi^i_0$ is a constant field 
configuration 
that minimises the effective potential,  
the second term in \variation\ vanishes. Since the first
term is just the mass matrix, we see that there is one  massless
particle for every  broken symmetry, which is precisely 
Goldstone's theorem [\GSW]. 

Differentiating \variation\ with respect to $\phi^i$, multiplying  by
$\omega^B f^j_B(\phi)\omega^C f^k_C(\phi)$ and taking 
$\phi^i=\phi^i_0$  we find  that
$$
{\partial^3 V(\phi_0)\over \partial \phi^i\partial\phi^j\partial\phi^k}
\omega^A f^i_A(\phi_0)\omega^B f^j_B(\phi_0)\omega^C f^k_C(\phi_0)
=0\ .
\eqn\threept
$$
Thus we see that at zero momentum the Goldstone three-point function
vanishes. Clearly we could keep differentiating and deduce that 
the Goldstone $N$-point function vanishes at zero
momentum for all $N$. 

In addition we could also repeatedly differentiate \eai\  and equation (2.2) 
with
respect to other fields $\psi$ that occur in the theory 
and we would conclude that their coupling to the 
Goldstone Bosons also vanishes at zero momentum. 

We will be interested in the Goldstone Bosons that occur in the presence of 
branes. Branes can be thought of a solitonic solutions to the classical 
equations of motion and before proceeding it will be helpful to remind 
the reader of how to include solitons in the quantum theory. 
Denoting the solitonic solution by 
$\phi_s^i$, we define the fluctuations about the soliton by 
$$
\phi^i = \phi^i_s +\phi_q^i\ ,
\eqn\phiq
$$
where $\phi_q^i$ is the quantum field. Next we expand the effective
action 
$$
\eqalign{
\Gamma[\phi_s+\phi_q]  &= \Gamma[\phi_s] 
+ \int d^Dx_1 {\delta\Gamma[\phi_s]\over\delta\phi^i(x_1)}\phi_q^i(x_1) \cr
&\qquad +{1\over2}\int d^Dx_1d^Dx_2
{\delta^2\Gamma[\phi_s]\over\delta\phi^i(_1)\delta\phi^j(x_2)}
\phi_q^i(x_1)\phi_q^j(x_2)+\dots\ .}
\eqn\qexpand
$$
Here the ellipsis indicates the higher order terms in $\phi^i_q$ which we
can ignore at lowest order in perturbation theory. The linear
term in $\phi^i_q$ vanishes since the soliton is a solution of the equations 
of motion. To proceed further we write $ \phi_q^i$ in terms of the 
complete set of 
solutions to the linearised equations in the presence of the soliton; 
that is the equations 
$$
\int d^D x_2 {\delta^2\Gamma[\phi_s]\over\delta\phi^i(x)\delta\phi^j(x_2)}
\eta^j (x_2) = 0\ .
\eqn\linear
$$
For a static soliton these equations become\foot{We have assumed for 
the sake of simplicity that the linearised equation can be written
in the form $-\partial_0^2\phi_q^i +\dots $ 
where the ellipses denote terms that don't
involve time derivatives.}
$$\int d^{(D-1)} {\vec x}_2
{\delta^2\Gamma[\phi_s]\over\delta\phi^i(x)\delta\phi^j(x_2)}
\eta^j({\vec x}_2) = -E^2 \eta^i\ ,
\eqn\slinear
$$
where we have taken $\eta^i(x)= e^{-iEt}\eta^i (\vec x)$. 
Labelling the solutions by the index I,  i.e. $\eta^i_I$,  we write 
$$
\phi^i_q(x)= \sum _I \eta^i_I(\vec x)  a^i_I(t) \ .
\eqn\modes
$$
The dynamics are now described by the variables $a^i_I(t)$ and  their action 
is found from the original functional integral by substituting the 
expression for $\phi^i_q$ above into
equation \phiq\
and keeping  terms second order in $a^i_I$.  

If the soliton solution depends on only some of the spatial coordinates, 
as is generally the case for branes,  then equation \linear\ is written 
differently. In particular, if the solitonic solution depends only
on the coordinates $y^\mu,\ \mu= p+1, \ldots ,D-1$ and we denote 
the remaining coordinates, including time, 
by $x^\mu, \ \mu=0, 1,2,\ldots , p$ then equation \linear\ takes the form 
$$
\int d^{(D-1-p)}\vec x_1 
{\delta^2\Gamma[\phi_s]\over\delta\phi^i(x)\delta\phi^j(x_1)}
\eta^j = k^2_{||} \eta^i\ ,
\eqn\branelinear
$$
where $k^2_{||}=-(k^{0})^2+(k^{1})^2+\ldots +(k^{p})^2$. The 
fluctuations are then expressed as 
$$
\phi^i_q= \sum _I \eta^i_I (y)  a^i_I(x)\ .
\eqn\braneexpand
$$

Finally, we can derive the analogue of Goldstone theorem for branes. 
Let us suppose that the theory  has a symmetry 
$\delta \phi^i= \omega^A f_A^i (\phi)$, where $\omega^A$ are constant 
parameters. 
As a result the effective action $\Gamma$ obeys the equation 
$$
\int d^D x_1 {\delta \Gamma\over\delta\phi^j(x_1)} \omega^A f_A^j (\phi)=0\ .
\eqn\symm
$$
Differentiating with respect to $\phi^i(x)$ and taking the 
fields to be evaluated at the soliton solution $\phi^i_s$ we deduce that 
$$
\int d^D x_1{\delta^2\Gamma[\phi_s]\over\delta\phi^i(x)\delta\phi^j(x_1)}
\omega^A f_A^j (\phi_s)=0 \ ,
\eqn\symmtwo
$$
where we have again dropped a linear term that vanishes on-shell.
If the soliton only depends on the coordinates $y^\mu$,  
as described above,  then 
this equation becomes \branelinear\ but with $k_{||}^2=0$ and 
$\eta^i=\omega^A f_A^i(\phi_s)$.  In other words, any symmetries of the
action which are broken by the soliton lead to  solutions of the
linearised field equation which are independent of the worldvolume
coordinates of the branes. 
Hence when expressing the fluctuation $\phi^i_q$ in terms of the 
complete set we find a terms of the form 
$\omega^A(x)  f_A^i (\phi_s (y))$ where $\omega^A(x)$ describes one 
massless particle corresponding for each of the symmetries broken 
by the soliton.  Thus to every symmetry broken by the solitonic solution 
we find a massless particle  propagating on the space where the 
soliton lives. 
In the case of branes this means propagating in the worldvolume of
the brane. The action of these particular  modes is deduced from the 
original action and takes the form 
$$
\int d^{p+1}x \partial_\mu \omega^A \partial^\mu \omega^B g_{AB}\ ,
\eqn\metric
$$
where $g_{AB}$ can be interpreted as a metric on moduli space 
and is determined by carrying out the $y$-integration. 
In the quantum theory it is this metric that 
enters into the norm of the states. We recognise these Goldstone modes 
as the collective coordinates discussed in the literature (for a
review see [\J]). 

An exception to Goldstone's theorem for solitons occurs if the metric 
$g_{AB}$ turns out not to be finite as a result of 
the $y$-integration.
To illustrate this point consider the standard non-linear sigma 
model based on the coset $G/H$. Such a theory is described by the 
group element $g\ \in \ G$ and is invariant under 
$g \to g_0gh$ where $g_0\ \in\ G$ is a rigid transformation and 
$h\ \in\ H$ is a local transformation. The action takes the form 
$$
\int d^Dx \sum _i ({\rm Tr} (g^{-1}dg X^i))^2\ ,
\eqn\nonlin
$$ 
where $X^i$ are the coset generators. 
Let us suppose we have a static solitonic solution then we are interested in 
modes where $g_0$ depends on $t$ i.e. $g_0(t)$. At infinity $g\to 1$ and 
these modes have the action 
$$
\int dt \int d^{D-1}{\vec x} \sum _i 
({\rm Tr} (g_0^{-1}(t)\dot g_0(t) X^i))^2 \ ,
\eqn\nonlinmodes
$$
in this region. Clearly, this diverges for those elements $g_0$ 
that belong to the coset.  

Of course  such a simple sigma model action 
does not admit solitonic solutions above two dimensions.  
Indeed while Golstone modes require a spontaneously broken continuous 
symmetry with a corresponding continuous family of vacua,  solitons
are often associated with discrete vacua. Therefore 
it might seem contradictory to discuss both broken symmetry and solitons 
in the same theory. However such theories do exist. For example, 
one can add  higher order
terms to the non-linear sigma model action \nonlin\ 
which enable the existence of solitons. A familiar example of this
type is  provided by the  Skyrme model in four dimensions. 
In these situations the quantum states are not normalisable 
and the corresponding modes must be dealt with differently. 

Let us now briefly review the 
Skyrme model [\S] in order to illustrate 
this point in detail and more importantly because, as we will see, 
this model is very  
analogous to the situation that occurs for branes. 
The action for the Skyrme model is given by
$$
S_{Skyrme}[U] = \int d^4x {\rm Tr}\left(
{f_\pi^2\over16}\partial_\mu U \partial^\mu U^{\dag}
+{1\over 32e^2}[(\partial_\mu U)U^{\dag} , 
(\partial_\nu U)U^{\dag}]^2\right)\ ,
\eqn\skyrme
$$
where $f_\pi$ and $e$ are constants and $U(x)$ is an element of $SU(2)$.
This action is clearly invariant under the group $SU_L(2)\times
SU_R(2)$ which acts as $U \rightarrow A_LUA^{\dag}_R$ where
$A_{L/R}\in SU_{L/R}(2)$. However
we must choose a vacuum configuration since any constant $U$ has zero
energy and can be used to define a vacuum state. Without loss of generality
we choose the vacuum to be $U=1$. Such a choice breaks the 
$SU_L(2)\times SU_R(2)$ symmetry to $SU_D(2)$ generated by taking
$A_L=A_R$. Thus there will be three Goldstone Bosons 
corresponding to
the broken symmetry. To exhibit these modes we write
$$
U(x) = e^{i\pi^i(x)T^i}\ ,
\eqn\pions
$$
where $T^i$, $i=1,2,3$ are the Pauli matrices and $\pi^i$ are the
Goldstone modes, i.e. the pions. It is clear from the action that
the pions are massless and, 
in the limit of zero momentum, there are no interactions of the
Goldstone modes among themselves.

As pointed out by Skyrme many years ago the inclusion of the higher 
derivative term in \skyrme\ allows for static soliton solutions of the
form 
$$
\pi^i = G(r){x^i\over r} \ ,
\eqn\skyrmion
$$
where $r = |\vec x|$ and we impose the boundary condition that 
$G(0)=\pi$ and $G(r)\rightarrow 0$ as $r\rightarrow\infty$. The
solutions to these equations have a finite energy, are topologically stable 
and  well-known as  ``Skyrmions''.
 
It is instructive to examine the Skyrmion  zero modes. 
First there are the  zero modes associated
with the breaking of translation invariance, which occur
for any finite energy soliton and are given by 
$f_\mu^i(\phi_s)=\partial_\mu \phi_s^i$.
In addition if $U_s$ is a Skyrmion then clearly
$A_LU_sA_R^{\dag}$ also solves the field equations for 
any constant choices of $A_L$ and $A_R$.
However let us promote $A_L$ and $A_R$ to 
time-dependent matrices. Substituting
$A_L(t)U_sA_R^{\dag}(t)$ into the action \skyrme\ we may 
determine an effective action for the modes $A_L$ and $A_R$. 
This leads to a rather complicated expression. However we can see that 
not all such modes have a finite kinetic energy. In particular
the leading order behaviour for large $r$ is
$$
\eqalign{
S_{eff}&=S_{Skyrme}[A_L(t) U_sA_R^{\dag}(t)]\cr
&= {\pi f_\pi^2\over 4}\int dt \int r^2 dr {\rm Tr}\left(
\left[\dot A_L\dot A_L^{\dag}+\dot A_R^{\dag}\dot A_R 
-\dot A_L^{\dag}A_LA_R^{\dag}\dot A_R
-\dot A_L\dot A_R^{\dag}A_R A_L^{\dag}\right]
\right)\cr 
&\qquad+ \ldots \ ,\cr}
\eqn\skyrmemodes
$$
where a dot denotes a time derivative and 
the ellipsis refers to terms with smaller powers of $r$ in the limit 
$r\rightarrow \infty$.
Thus we see that unless $A_L=A_R$ the spatial integral will diverge.
Indeed the dominant terms at large $r$ are simply those of
a non-linear realisation as discuss above.
Furthermore one can check that if $A_L=A_R$ then the sub-leading terms
in \skyrmemodes\ are finite [\ANW]. 
Therefore only the zero-modes corresponding to $SU_D(2)$
can be associated with the motion of the soliton.  The broken symmetries
generate new solutions to the field equations which cannot be simply 
interpreted as dynamical states of the Skyrmion.

This is clearly a general phenomenon. 
If the field equations  of a theory are invariant under a continuous
symmetry group then any soliton will come with zero modes:
i.e. by acting on a soliton with the symmetry generators we will
obtain a new solution to the field equations (unless the soliton itself
is invariant under the symmetry in question). 
However it is important to distinguish between what might be called genuine
zero-modes of a soliton and non-genuine zero-modes. 
Genuine zero-modes  have a finite
kinetic term. Thus these modes appear in the low energy motion of
the soliton in the form of collective coordinates. However there may
also be zero-modes which do not have a finite kinetic term and so
cannot be interpreted as collective coordinates. We call these non-genuine
zero-modes.

If there are non-genuine zero modes
then we will find solutions to the field equations which cannot be
associated with the motion of the soliton. Acting 
on the vacuum  with the broken symmetry generators 
produces a Goldstone state.
Therefore we intuitively expect that acting on a soliton with a 
broken symmetry generator
(i.e. non-genuine zero-mode) produces a soliton/Goldstone state. We
will show in the next section that this is in fact 
just a soliton/Goldstone scattering
state at zero momentum.


\chapter{Goldstone/Soliton Interactions at Zero Momentum}

In this section we will now consider the interactions of Goldstone modes
with solitons in a general quantum theory. We will be particularly
interested in the scattering of Goldstone modes off a soliton at zero 
momentum.  
In the beginning of section two we explained why Goldstone particles 
had only derivative interaction with themselves and all other particles. 
This argument was based on the effective action viewpoint. 
Since the soliton is not represented by its own field,
that derivation does not apply to the scattering of
Goldstone particles off solitons. In fact we will see that the 
result does not hold in general and in some cases one indeed finds
non-derivative Goldstone soliton interactions. 

We take the approach to solitons reviewed in [\J]. 
We imagine that the  theory 
has a symmetry which is spontaneously broken leading to 
a corresponding Goldstone 
particle. These modes are included in the generic symbol $\phi^i$. 
We suppose that the theory also possess a stable classical soliton 
solution which we denote by $\phi^i_s$. 
Corresponding to this classical solution 
there is an associated quantum state of momentum $\vec {p}$ denoted by 
$|\vec {p}>_s$.
We normalise the Lagrangian so that the fields occur in the
combination $g\phi^i$, where $g$ plays the role of a coupling
constant and we include an overall factor of $g^{-2}$ in front of the
action
$$
S = {1\over g^2} \int d^Dx {\cal L}(g\phi^i)\ .
\eqn\action
$$
Solutions to the equations of motion, in particular the solitons 
$\phi_s^i$, are then of order $g^{-1}$ and the mass of the soliton 
states are  of order $g^{-2}$.

We must also consider states which contain the soliton and the 
elementary particles that occur in the sector of the theory that has no 
solitons. The latter include the Goldstone particles of the theory. The 
simplest examples 
are the states $|\vec {p};\vec{k}_1,\vec{k}_2,...>_s$ where $\vec {p}$ is the 
momentum of 
the soliton and $\vec{k}_1,\vec{k}_2,...$ are the momenta of the 
Goldstone particles or other fundamental particles. 
The connected components of the matrix elements between two states 
are assumed to be of order $g^{n-1}$  
where $n$ is the number of elementary particles involved. 
We also assume that
the soliton is stable so that the matrix element between a state in
the soliton section and a state in the vacuum sector vanishes.
With these assumptions one can show that, to lowest order in $g$,
$$
\phi^i(x) = {1\over (2\pi)^{D-1}}\int d^{D-1} (p-q)
e^{i(\vec{p}-\vec{q})\cdot\vec{x}}{}_s 
<\vec {p}|\phi^i(0)|\vec {q}>_s\ ,
\eqn\solone
$$
solves the field equations and we therefore identify it with the
soliton solution $\phi_s^i(x)$. Furthermore, if we introduce  
the Fourier transform of the matrix element with a Goldstone state 
$$
\eta^i_k (\vec {x})= {1\over (2\pi)^{D-1}}
\int d^{D-1}(p-q) e^{i(\vec {p}-\vec{q})\cdot \vec{x}}
{}_s <\vec {p}|\phi^i(0)|\vec {q};\vec{k}>_s 
\ ,
\eqn\soloneone
$$
then one can show that $e^{-iE_{k}t}\eta^i_k(\vec{x})$
solves the linearised equation of motion \linear\ 
in the background of the soliton.

We now will derive the interaction of the soliton with the 
Goldstone particles at zero momentum. We begin with  the relation 
$$
[Q_A, \phi^i]=i\delta_A \phi^i\ ,
\eqn\varone
$$
where $Q_A$ is one of the symmetry generators of the theory 
that is spontaneously broken. Taking the scalar product of the left hand side 
with 
solitonic states and writing the charge as the integral of its current 
$j^\mu_A$ we  find that 
$$
\int d^{D-1} \vec {x} \ {}_s <\vec {p}|j^0_A(\vec {x},t)\phi^i|\vec {q}>_s
-{}_s <\vec {p}|\phi^i j^0_A(\vec {x},t)|\vec {q}> 
= i\ {}_s<\vec {p}|\delta_A\phi^i|\vec {q}>_s\ .
\eqn\soltwo
$$
Using translational invariance we choose to evaluate 
$\phi^i$ at the origin of the coordinates, i.e. $\phi^i = \phi^i(0)$. 
Sandwiching 
with a complete set of states $|n>$, the left-hand side of \soltwo\ becomes 
$$
\sum_n \int d^{D-1} {\vec x} \ 
{}_s <\vec {p}|j^0_A(\vec {x},t)|n><n|\phi^i|\vec {q}>_s
-{}_s <\vec {p}|\phi^i |n><n|j^0_A(\vec {x},t)|\vec {q}>_s
\eqn\solthree
$$
In the complete set, the only states that contribute are the 
one soliton states.  Goldstone's theorem
asserts that acting on the vacuum with a broken $j^0_A$ creates a
massless particle which  carries  the same quantum numbers as the
current. In the soliton sector of the theory we can use cluster
decomposition to argue that a broken $j^0_A$ also creates a Goldstone
mode in the soliton background, since far from the soliton core it
must act in the same way as it does in the vacuum.
Consequently, for a broken symmetry $Q_A$, the intermediate matrix
element
${}_s<\vec {p}|j^0_A(\vec {x},t)|n>$ will be non-vanishing only if 
the states $|n>$ 
are a one soliton state with an appropriate Goldstone particle, that 
is the state denoted $|\vec {p'};\vec{k'}>_s$.  
In some cases, for example if all the solitons carry the same
quantum numbers,  this may also be seen as a consequence of the 
conservation of the  quantum number associated to $j_A^\mu$.
Using translational
symmetry, that is
$$
{}_s <\vec {p}|A(\vec{x},t)|n>= {}_s <\vec {p}|A(0,0)|n>e^{-i(E_p-E_{p_n})t}
e^{i(\vec {p}- \vec {p}_n)\cdot \vec {x}}\ ,
\eqn\solfour
$$
we may insert the Goldstone/soliton states
and carry out the $\vec {x}$ integration.  We 
then find that the left hand side  of equation \soltwo\ becomes 
$$
\eqalign{
(2&\pi)^{D-1}  \int {d^{D-1}\vec {p'}\over 2E_{p'}}
\int {d^{D-1} \vec {k'}\over 2E_{k'}}\left\{ \right.\cr
&\left.\delta^{D-1}(\vec {p}-\vec {p'}-\vec{k'})
e^{-i(E_p-E_{p'}-E_{k'})t}
{}_s <\vec {p}|j^0_A(0,0)|\vec {p'};\vec{k'}>_s 
{}_s<\vec {p'};\vec{k'}|\phi^i|\vec {q}>_s \right.\cr
&\left.- \delta ^{D-1}(\vec {p'}-\vec {k'}-\vec {q})
e^{i(E_q-E_{p'}-E_{k'})t}
{}_s<\vec {p}|\phi^i|\vec {p'};\vec{k'}>_s 
{}_s <\vec {p'};\vec{k}'|j^0_A(0,0)|\vec {q}>_s \right\} \ .}
\eqn\bigeq
$$
We will evaluate the 
above quantity to lowest order in the coupling $g$. 
At this order the matrix element 
${}_s<\vec {p}|j^0_A(0,0)|\vec {p'};\vec{k'}>_s$ 
takes the form of a disconnected diagram, that is 
$$
{}_s<\vec {p}|j^0(0,0)_A|\vec {p'};\vec{k'}>_s= {}_s<\vec {p}|\vec
{p'}>_s 
<0|j^0_A(0,0)|\vec{k'}>\ .
\eqn\disscon
$$
This quantity is of order $g^0$. 
Let us assume that the current $j^\mu_A$ carries no Lorentz index 
other than $\mu$. Imposing  Lorentz invariance we deduce that 
$<0|j^\mu_A(0)|\vec{k'}> =iF_A k'^\mu$ where $F_A$ is constant 
of order $g^0$ which we may take to be real.  Hence we obtain 
$$
{}_s<\vec {p}|j^\mu_A (0,0)|\vec {p'};\vec{k'}>_s
= 2iF_A k'^\mu E_{p} \delta^{D-1}(\vec {p}-\vec {p'}) \ .
\eqn\discontwo
$$
Substituting in this  expression, and using the on-shell
relation $E_{k'} = |{\vec k'}|$ for massless fields, we arrive at 
$$
i (2\pi)^{D-1} F_A \lim_{k'\to 0}
\left( {}_s<\vec {p}; \vec{k'}|\phi^i|\vec {q}>_s
+{}_s<\vec {p}|\phi^i|\vec {q};\vec{k'}>_s \right)
= i\ {}_s <\vec {p}|\delta_A \phi^i|\vec {q} >_s\ .
\eqn\solfive
$$

We recall that 
the generator $Q_A$ is one of the symmetry generators that is spontaneously 
broken in the vacuum sector and as a consequence 
leads to Goldstone particles. We have assumed that the   
soliton also breaks this symmetry, i.e. $Q_A|\vec {p}>_s\not= 0$, and as a 
result ${}_s <\vec {p}|\delta_A \phi^i|\vec {q} >_s$ is non-vanishing. 
In fact, to 
lowest order,   ${}_s <\vec {p}|\delta_A \phi^i|\vec {q} >_s$, 
or more precisely its
Fourier transform,  is a solution to the
linearised equation of motion about the soliton background.
Hence we conclude that the matrix element 
${}_s<\vec {p}, \vec {k}'|\phi^i|\vec {q}>_s$ 
at zero Goldstone momentum is related to the variation of the
soliton under a broken symmetry. 

The above argument parallels the classic proof of Golstone's theorem [\GSW]. 
In this case one begins with the same relationship of equation \soltwo\
but takes the matrix element to be between the vacuum states rather than the 
solitonic states. One then proves that there exist massless particles 
with the same quantum numbers as the broken generators.  This follows as one 
shows that there must exist a massless particle $|\vec{k}>$ 
in the complete set 
such that $<0|j^0|\vec{k}>$ and $<\vec{k}|\phi^i|0>$ are non-vanishing. 

In fact by analogy with the original proof of Goldstone's theorem
we can prove a slightly stronger result by considering 
the  time derivative of the relation $[Q,\phi^i]=\delta \phi^i$. 
Writing $\partial _0  Q$ as an integral over $\partial _0 j^0$ we can  
express the volume 
integral  as a surface integral at infinity. Using the fact that 
operators at space-like separation commute,  we conclude that 
${}_s<\vec {p}|\delta \phi^i|\vec {q}>_s$ is independent of time. 
On the other hand if we evaluate \bigeq\ then the requirement of 
time independence implies the massless on-shell condition 
$\lim_{\vec{k}\rightarrow 0} E_{\vec {k}} =0$. 
Therefore we learn that the intermediate particles must be
massless  modes.

For completeness we consider the analogous proof if we 
choose a symmetry $Q_A$
that is unbroken in the vacuum. 
In this case it is clear that the internal symmetry
will lead to a moduli space of solitons. In the quantum theory 
there will be discrete orthonormal states, 
represented by wave functions on this
moduli space, which are interpreted as distinct soliton
states. Although all these soliton
states carry the same soliton number, they may carry 
different quantum numbers such as spin and, in the case of the Skyrme
model, isopin [\ANW]. Therefore we introduce the indices  
$\alpha, \beta,...$ and label the soliton states as
$|\vec{p},\alpha>$. Note that this does not substantially affect
the previous discussion since, upon using \disscon, we would learn 
that $F_A$ 
is replaced by matrix $F_A^{\alpha\beta}$ which is diagonal. In which
case the previous discussion applies for each type of soliton separately.

Returning to the argument we now see that there is no
Goldstone particle  in the intermediate state, at least not at lowest 
order in
$g$, since there is no corresponding Goldstone mode in the vacuum
sector. Rather we would find the intermediate state is another soliton
$|\vec{p'},\gamma>_s$. 
Assuming that the charge $Q_A$ and solitons are scalars,  
Lorentz invariance  now restricts the 
soliton/current/soliton matrix element to be of the form
$$
{}_s<\vec{p},\alpha|j^\mu_A(0,0)|\vec{p'},\beta>_s = G_A^{\alpha\beta}
(p+p')^\mu\delta^{D-1}(\vec{p}-\vec{p'})\ ,
\eqn\solsix
$$
with $G_A^{\alpha\beta}$ a constant Hermitian matrix of order $g^{0}$.
Note that  Lorentz invariance alone also allows for a term proportional to
$(\vec{p}-\vec{p'})^\mu$ but this term vanishes if $j^\mu_A$ is
conserved. In addition elements of $G_A^{\alpha\beta}$
where the solitons  labelled by $\alpha$ and
$\beta$ have the different  masses must also vanish. 
If we now continue with the argument as above  we 
conclude that
$$\eqalign{
(2\pi)^{D-1}\sum_\gamma \left(
{}_s<\vec {p},\alpha|\phi^i|\vec {q},\gamma>_sG_A^{\gamma\beta}
\right.
&\left.
-G_A^{\alpha\gamma}{}_s<\vec {p},\gamma|\phi^i|\vec {q},\beta>_s \right)\cr
&=  i\ {}_s<\vec {p},\alpha|\delta_A \phi^i|\vec {q},\beta >_s\ .}
\eqn\solseven
$$
In other words we simply find that the unbroken symmetries are
linearly realised on the solitonic states. In terms of matrices
we see that the variation of 
${}_s<\vec {p},\alpha|\delta_A \phi^i|\vec {q},\beta >_s$ is given by 
its commutator with  $G_A^{\alpha\beta}$.

Our last step is to determine the scatting amplitudes from the
matrix elements 
${}_s<\vec{p}|\phi^i(l)|\vec{q}>_s$ and 
${}_s<\vec{p}|\phi^i(l)|\vec{q};k>_s$ where $l$ is the Goldstone
momentum. For simplicity we will drop any reference to the
soliton indices $\alpha,\beta,...$ since it will be clear that they  
will not
affect the main point of the discussion.
As is the case in standard
scattering theory, if we view the states $|\vec{p}>_s, \ 
|\vec{p};\vec{k'}>_s$ 
and ${}_s<\vec{p}|, \ {}_s<\vec{p};\vec{k}|$ 
as incoming and outgoing respectively, then 
we may use the LSZ reduction formula to relate these elements to
soliton/soliton/Goldstone and soliton/soliton/Goldstone/Goldstone 
scattering. 
These matrix elements can be determined from the matrix
elements dicussed above by a Fourier transform and use of the
formula \solfour
$$\eqalign{
{}_s<\vec{p}|\phi^i(l)|\vec{q}>_s &= 
\int d^4x e^{-il\cdot x}e^{i(p-q)\cdot x}
{}_s<\vec{p}|\phi^i(0)|\vec{q}>_s \cr
{}_s<\vec{p}|\phi^i(l)|\vec{q};\vec{k}>_s &= 
\int d^4x e^{-il\cdot x}
e^{i(p-q-k)\cdot x}
{}_s<\vec{p}|\phi^i(0)|\vec{q};\vec{k}>_s \ .\cr 
}
\eqn\scatone
$$
In turn we may use \solone\ and \soloneone\ to express the right hand side of
\scatone\ in terms of the Fourier transforms of the soliton solution
and the solution to the linearised equation respectively.
The first of these equations simply involves the soliton solution $\phi_s^i$
whereas the second equation involves solutions to the  linearised
field equation in the background of the soliton. 
However here we are interested in the scattering at zero Goldstone momentum 
and from \solfive\ we learn that the corresponding matrix elements
are obtained by acting with the symmetry variation 
corresponding to the  broken symmetry  on the soliton solution. 
We therefore find, in the zero Goldstone momentum limit,
$$
{}_s<\vec{p}|\phi^i(l)|\vec{q}>_s = 
(2\pi)^D\delta(l^0)\delta^{D-1}(\vec{p}-\vec{q}-\vec{l})
\int d^{D-1}y e^{i(\vec{p}-\vec{q})\cdot\vec{y}}\phi^i_s(\vec{y}) \ ,
\eqn\scattwo
$$
and
$$\eqalign{
\lim_{k\rightarrow 0}\left({}_s<\vec{p}|\phi^i(l)|\vec{q};\vec{k}>_s
+{}_s<\vec{p};\vec{k}|\phi^i(l)|\vec{q}>_s\right)
&= \cr
{2\pi\over F_A}\delta(l^0)\delta^{D-1}(\vec{p}-\vec{q}-\vec{l})
&\int d^{D-1}ye^{i(\vec{p}-\vec{q})\cdot\vec{y}}
\delta_A\phi^i_s(\vec{y})\ ,\cr }
\eqn\scatthree
$$
where $\delta_A$ is a broken symmetry variation.
In deriving \scattwo\ we have used the fact that 
$E_{\vec{p}}-E_{\vec{q}}=0+{\cal O}(g^2)$ 
for any two solitons with the same rest mass.

The LSZ formula  asserts
that the disconnected part of the scattering amplitude is simply
the residue of the $1/l^2$ term in the matrix element, up to a
normalisation constant. Imposing the delta function constraint $l^0=0$,
the three-point  scattering
amplitude between a Goldstone mode and two solitons comes from the 
$1/r^{D-3}$ term (${\rm ln} r$ term in $D=3$) in
$\phi_s^i(r)$ 
where $r=|\vec{y}|$.
Similarly the 
scattering amplitude of two Goldstone particles and two solitons, at
zero Goldstone momentum, is given by the coefficient of the $1/r^{D-3}$
term  (${\rm ln} r$ term in $D=3$) 
in the solution to the linearised equation of motion \linear\
$$
\delta_A\phi_s^i =  f_A^i(\phi_s)\ ,
\eqn\scatfour
$$
generated by a broken symmetry.

Let us return to our example of the Skyrme model in four
dimensions. Here one finds that,
under an infinitesimal $SU_L(2)\times SU_R(2)$ symmetry, the field
transforms as
$$
\delta \pi^i = (g^i_L - g_R^i)|\pi|{\rm cot}|\pi|
-\pi^i{(g^j_L-g^j_R)\pi^j\over |\pi|^2}(|\pi|{\rm cot}|\pi|-1)
-\epsilon^{ijk}(g_L^j+g_R^j)\pi^k\ ,
\eqn\variation
$$
where $A_{L/R}= e^{ig^i_{L/R}T^i}$ and $|\pi|^2 = \pi^i\pi^i$. 
The broken generators are $g^i_L=-g_R^i$ and we have seen that these 
give rise to non-genuine zero-modes. Substituting the Skyrmion soliton 
\skyrmion\
into \scatthree\ and taking $g_L^i=-g_R^i$ we obtain 
$$
\delta\pi^i = 2g_L^j\left(\delta_j^i G{\rm cot}G
-{x^ix^j\over r^2}(G{\rm cot}G-1) \right) \ ,
\eqn\skyrmscat
$$
which, as we have argued above, solves the linearised field equation \linear.
However  a Skyrmion behaves like 
$\pi_s^i \sim G(r)= {\cal O}(1/ r^{2})$ as $r\rightarrow \infty$. 
Thus the scattering
solutions \skyrmscat\ behave like 
$\delta\pi^i_s=2g^j_L+{\cal O}(1/r^2)$ at infinity.
Hence we conclude that in this model there are  vanishing  
three-point and four-point scattering amplitudes
between the Goldstone modes and the Skyrmion at zero momentum. Indeed
this is the case [\MK].  

On the other hand, 
for any  soliton that carries electric or magnetic charge,
the gauge field  must have a non-vanishing $1/r^{D-3}$ term at infinity as
a consequence of  Gauss' law.  
We therefore see that these
states have non-derivative soliton/soliton/gauge field couplings.
Indeed this is just the familiar minimal coupling of
a field to an electromagnetic potential. 
In the case of branes the low energy $U(1)$ gauge field arises
as a Goldstone mode. In this way we see how the charged solitons 
can couple minimally to the gauge field. In the next section we will discuss
in more detail other examples of
non-derivative Goldstone/soliton scattering in the effective theory
of branes.

\chapter{Applications to Branes}

In this section we wish to apply the general theory that we discussed
above to the specific case of a wrapped M-fivebrane.  
Let us therefore begin  with a discussion of
M-fivebrane dynamics
and in particular its relation to quantum gauge theory.

\section{Review of the M-fivebrane and Gauge Dynamics}

For obvious reasons we are
most interested in  M-fivebranes that have only four macroscopic
spacetime dimensions.  There are essentially two ways to realise  this. 
The first is
to consider intersecting M-fivebranes. In this case the
four-dimensional
spacetime is the intersection where the worldvolume fields become
localised. Curiously, at least within the low energy approximation to
the M-fivebrane, a smooth intersection is equivalent to a single
M-fivebrane whose worldvolume appears wrapped on a calibrated (and
in general non-compact) surface $\Sigma$
[\W,\HLWthree,\GP,\GLWone,\QMW]. 
Note that in
this case the bulk spacetime need not have any non-trivial topology of
its own, one is merely choosing a non-trivial  embedding 
the M-fivebrane into the bulk. On the worldvolume of the M-fivebrane
this intersection appears as a solitonic solution with only scalars
active.   
The other possibility is simply to take the bulk eleven-dimensional
spacetime to be of the form ${\bf R^4} \times {\cal M}$, where 
${\cal M}$ is a non-trivial seven-manifold which has a topologically
non-trivial
two-cycle $\Sigma$  over which we may wrap the M-fivebrane. Therefore both
these constructions are similar in that they involve 
M-fivebranes which are in some sense ``wrapped'' 
over a two-dimensional surface $\Sigma$. 
However there is at least
one crucial difference. In the first case, where the intersection is
realised as a non-trivial embedding, there will generically be
moduli of this embedding which will show up as massless fields on
the M-fivebrane. On the worldvolume of the M-fivebrane the
intersection appears as a soliton solution and 
these scalars may  be thought of as Goldstone fields
for broken translations. In the second case 
the M-fivebrane will wrap the two-cycle in such a
way as to minimise its volume. This will be fixed by the properties
of the bulk spacetime and therefore  we don't expect 
any massless scalar moduli on the M-fivebrane.

The simplest example is the case of two static M-fivebranes intersecting 
over a common three-dimensional space,  preserving eight 
supersymmetries. 
The M-fivebrane worldvolume description of this configuration is 
that of just a single M-fivebrane with  
two of its worldvolume dimensions wrapped over a non-compact  
Riemann surface [\W,\HLW]. The effective action for the
zero-modes of the corresponding soliton can be constructed as 
outlined in section two and
agrees  precisely with the Seiberg-Witten low energy effective theory for 
$N=2$ Yang-Mills theory [\HLW]. 

The appearance of the Seiberg-Witten effective action  
is not a coincidence but rather a prediction of the duality
between M-theory and type IIA string theory [\W]. By compactifying on 
a circle the intersecting M-fivebranes are interpreted as intersecting
D-fourbranes and NS-fivebranes in type IIA string theory. The
description of D-branes in terms of open strings can be used to  show 
that the brane
dynamics are given by $N=2$ Yang-Mills theory [\W]. Therefore one expects 
that the low energy dynamics
of the M-fivebrane should precisely reproduce the quantum low energy
effective action of $N=2$ Yang-Mills gauge theory.  This 
example illustrates a deep relationship between the low energy
dynamics of the M-fivebrane and quantum gauge field theories.
It is also natural to wonder to what extent this relationship
can be applied to more realistic gauge theories.

One of the central problems in string or M-theory is how to relate its 
effects to the observable world, 
in other words the standard model, in a convincing way. It could be that one 
has to understand much more about M-theory than we currently do 
in order to achieve this.  However, 
it is also possible that we are at least in a position to  
find some convincing signs. There is a very large literature on relating 
supersymmetric theories, then string theories and more recently 
theories including branes to the world we observe.
Almost all of these approaches have tried to 
construct not the standard model directly, but either 
the minimal supersymmetric extension of the standard  model 
or some supersymmetric grand unified extension. As such these papers, 
including much of the recent work, is based on trying to find a 
realistic model which possesses hidden sectors or soft breaking terms.  

In a previous  paper [\LW] we discussed a more direct approach.   
The standard model involves the gauge groups $SU(2)\times U(1)$ 
broken to a $U(1)$ and a  confined $SU(3)$. It also 
has chiral Fermions and no supersymmetry. These rather generic 
features arise in a natural way from 
intersecting branes.  
Hence the paper [\LW] addressed the question
as to whether or not one could, in principle, find a  
wrapping of the M-fivebrane that lead directly to the standard model. 
That is to say to find a wrapping of the M-fivebrane whose effective
theory is the same as the effective theory of the standard model.

One immediate problem which one faces is that the states that arise on
D-branes are only charged with respect to two
gauge groups, one for each end point of an open string. However quarks
carry charges under all of the three simple factors of 
$SU(3)\times SU(2)\times U(1)$. In fact D-branes give rise to
$U(N)$ gauge groups and  the over-all $U(1)$ generally decouples.
Therefore, as a first step in this direct approach, in [\LW]
M-fivebrane  configurations with gauge group 
$SU(3)\times SU(2)\times U(1)$ and containing states
charged under all three groups were found. Furthermore 
the assignments of this ``hypercharge'' were
derived from the brane physics and lead to a  realistic spectrum. 
Recently  the  authors of [\AKT,\AIQU] have taken other ``bottom-up''
approaches and [\AIQU] discussed
mechanisms for hypercharge in perturbative type IIB string theory
with similar features. It is possible that these mechanisms are
related by T-duality.

More specifically, in [\LW] it was shown that
M-fivebrane configurations exist with $N=2$ supersymmetry and the  
gauge group $SU(N_1)\times SU(N_2)\times ... \times SU(N_k)\times U(1)$. 
Furthermore these
models possess hypermultiplet states in the $({\bf N}_a, {\bf N}_b)$ of 
$SU(N_a)\times SU(N_b)$ for each pair of simple factors labelled by 
$a,b=1,2,...,k$. 
The $U(1)$ charge of these multiplets is
determined to be $\pm(N_a^{-1}-N_b^{-1})$. In this
way a natural toy model can be  constructed 
with the gauge group  $SU(3)\times SU(2)\times U(1)$  
and ``quark'' multiplets in the representation 
$({\bf 3},{\bf 2}, \pm{1\over6})$. 

It is also possible to add
$n$ D-sixbranes into these models along the lines of [\W]. Each
D-sixbrane introduces $k$ hypermultiplets in the ${\bf N}_a$ of $SU(N_a)$ 
which carry a $U(1)$ charge  $\pm N^{-1}_a$.
These multiplets are generically massive, however it is possible to
tune some of them to be massless. Including a single D-sixbrane into
our toy model we may 
obtain a massless ``lepton'' multiplet in the 
$({\bf 2}, \pm{1\over2})$ of $SU(2)\times U(1)$\foot{We thank 
A. Uranga for pointing this out to us.} although it must
also come with a massive mutliplet in the 
$({\bf 3}, \pm{1\over3})$ of $SU(3)\times U(1)$.

Of course such a toy model is still far from realistic. For instance it is
non-chiral, has $N=2$ supersymmetry and it is not clear how 
three generations can be incorporated. 
Nevertheless we find it encouraging that
a similar structure to what is found in the standard model arises so
readily from branes. In [\LW] various steps were discussed that
would lead to a more realistic model. For example the supersymmetry
can be broken and the massless scalar modes removed by wrapping the
M-fivebrane over a non-supersymmetric two-cycle in ${\cal M}$.

\section{Goldstone Neutrinos}

An M-fivebrane in eleven dimensions can be thought of as a solitonic 
solution of eleven-dimensional supergravity,  or at a more fundamental level,
of the underlying M-theory. 
It breaks translational invariance and 
half of the supersymmetries of M-theory. As expected 
we find that the dynamics of the fivebrane includes five Goldstone Bosons
and sixteen Goldstone Fermions. There is also a self-dual three-form tensor
field which arises from the breaking of certain automorphism
symmetries of the M-theory algebra [\West] and is the field strength
for an Abelian two-form gauge field. In other words, all the massless modes 
of a brane in M-theory arise via non-linear realisations and therefore
only  have derivative couplings [\West].

For wrapped  M-fivebranes the resulting massless fields 
in the low energy effective action can therefore  be interpreted
as Goldstone particles which are confined to the four-dimensional 
worldvolume, in accordance with section two. 
In particular we expect Goldstone 
Fermions from the breaking of
the worldvolume supersymmetries. Furthermore these  Goldstinos 
cannot be absorbed by the gravitinos associated 
with the background spacetime as they are confined to the worldvolume of the 
self intersection. Put another way, they are not absorbed from 
the M-fivebrane viewpoint as this theory does not involve a 
dynamical gravitino. In addition there will generically  
be four-dimensional Abelian
vector modes which arise from the two-form gauge field. 
All these degrees of freedom can be described by an effective action 
that, at least in principle, can  be deduced from the dynamics of the 
M-fivebrane.  

Note that, since the  two-form arises as a 
Goldstone field on the M-fivebrane, it only has derivative 
interactions with itself and the other fields. That is to
say it only appears in the equations of motion through its three-form 
field strength. Therefore any vector gauge fields that 
appear on the four-dimensional worldvolume do so 
through their  field strengths. Hence all the fields
that occur on the M-fivebrane are neutral under these $U(1)$'s.
It follows that any charged states which arise in the dynamics of
the M-fivebrane worldvolume  must arise a solitons.
One can think of these solitons as M-twobranes, 
whose boundaries are wrapped over one-cycles in $\Sigma$ [\W]. In the dual
type IIA limit, obtained by compactification to ten dimensions,
these vectors arise from the open strings ending on D-fourbranes.
Some of these solitons have been studied in [\HLWone,\GLW,\LWm,\G]. 
Their charge arises from the fact that M-twobranes 
couple directly to the self-dual gauge field of the M-fivebrane
[\St,\T].   In particular for the $N=2$ configurations discussed 
above, the states which
correspond to the $W^\pm$ and monopoles
can be found as soliton solutions on the M-fivebrane worldvolume
[\GLW,\LWm,\G].

This picture of charged states appearing as solitons is 
analogous to the situation with branes and Ramond-Ramond charges that 
emerged after the U-duality conjecture [\HT]. U-duality 
necessarily implies that there are states in string theory 
which are charged with respect to the Ramond-Ramond
gauge fields. However it is well-known that no state in perturbative
string theory can carry  Ramond-Ramond charge, since these 
fields appear only through their field strengths. 
From the point of view of the low energy supergravity theory however 
one can readily find such states in the form of $p$-brane solitons
[\HS].  It is only
through the inclusion of D-branes [\P] that such charged states are 
identifiable in the fundamental theory.

In the models discussed in [\LW] the Goldstone 
particles resulting from the wrapping of the M-fivebrane include Goldstone 
Fermions from the breaking of supersymmetry and
vector particles corresponding to the unbroken $U(1)$'s. 
Since none of the massless fields on an M-fivebrane can carry the
charge of the gauge fields, these  
Goldstone Fermions are neutral and could  potentially 
be identified with neutrinos.  
Furthermore, as we have discussed above, in
such a scenario any states which are charged under the low energy  
$U(1)$, such as the electron, can only arise as solitons of the M-fivebrane 
equations of motion. 

It was noted [\LW] that such a brane world model 
provides a natural setting for 
the old idea of Volkov and Akulov to view the neutrino as a Goldstino. 
Indeed in the simplest example  of an M-fivebrane
wrapped on a Riemann surface the Fermions
appearing in the low energy effective action 
are Goldstinos.  However  their low energy dynamics is indistinguishable from
that of a  theory where they have non-derivative couplings, namely $N=2$
Yang-Mills.  Furthermore 
M-theory/type IIA duality strongly suggests that in the full 
M-fivebrane dynamics the Goldstinos must have have non-derivative
interactions with the electrically  charged states.  In other words
we see that a soliton/Goldstino non-derivative 
interaction must exist. Given our discussion in section three we are
now in a position to see how this is possible.

In fact the Skryme model discussed above 
is quite analogous to the situation 
with  the effective theory of the wrapped M-fivebrane. 
The Skryme model was constructed
as a description of the low energy physics of the strong interactions. 
The pions arise as the 
Goldstone modes for an $SU_L(2)\times SU_R(2)$ symmetry broken to $SU_D(2)$. 
Skyrme's insight was to suppose that nucleons could also appear
in the low energy description as solitons. 
Of course, we now know that the fundamental theory of 
strong interaction is QCD and presumably an action similar to the  
Skyrme model is derivable from it in a suitable low energy limit.  

Thus QCD plays an analogous role to M-theory 
and the Skyrme model is the counter 
part of the effective theory of the wrapped M-fivebrane.
The fundamental fields on the M-fivebrane arise as Goldstone modes,
just as the pions do in  the Skyrme model.  
In addition the nucleons are realised as solitons in 
the Skryme model and  in the  
effective theory of the wrapped fivebrane all charged states must
arise a solitons. Thus the process we wish to 
study is the scattering of the solitons and the fundamental 
particles or Goldstone modes in each theory. 
That is the scattering of the nucleons with pions in the Skyrme model and 
neutrinos with electrons and $W^\pm$'s
in the effective theory of the wrapped fivebrane. 

These two models have another point in common. In the low energy limit, 
the lowest order terms for an M-fivebrane wrapped on
a Riemann surface are known to be given by the 
Seiberg-Witten effective action
[\HLW]. However, 
the Seiberg-Witten effective action does not support a finite energy
soliton configuration that represents either
a monopole or $W^\pm$ [\LWm], as one expects from
Derricks theorem. However it can be argued that suitable
corrections to the Seiberg-Witten dynamics do arise from the 
M-fivebrane that will support such soliton solutions [\LWm]. Similarly
in the
Skyrme model the higher derivative term must be added in order to
enable the existence of the solitonic nucleons. 

However there is an important difference between the two models. 
Namely the soliton solutions on
the M-fivebrane that describe charged states behave as $1/r$ at
infinity [\LWm] and not $1/r^2$ as the Skyrmion does. 
Therefore, according to our argument in
section three, there will be non-derivative scattering of the solitons 
with the Goldstone particles. Although we only considered bosonic
internal symmetries in section three 
we expect an analogous result will be apply for 
supersymmetries and their corresponding Goldstinos. In these cases one
would have to repreat this analysis including spinorial and Fermionic
charges.
Indeed, as we mentioned in the
introduction, in the example of two intersecting M-fivebranes with 
four-dimensional $N=2$ supersymmetry,  duality implies that these
non-derivative interactions 
must exist on the M-fivebrane since they exist in the gauge theory
of the D-fourbranes in type IIA string theory. In this sense the
argument in section three provides a test of M-theory/type IIA duality
and the relation of the M-fivebrane to quantum gauge theory.

Returning to 
realistic scenarios the four-dimensional M-fivebrane vacuum
itself breaks rigid supersymmetry by wrapping over a
non-supersymmetric two-cycle of ${\cal M}$, i.e. the four-dimensional
vacuum of the M-fivebrane spontaneously breaks rigid supersymmetry. 
This leads to massless
and chargeless  Goldstinos in the four-dimensional worldvolume. In
addition any electrically charged states arise as soliton solutions
to the M-fivebrane worldvolume theory and these necessarily 
also break the supersymmetry of the M-fivebrane equations of motion. 
Hence it follows that, in the effective theory 
of the wrapped fivebrane, the spin 1/2 Goldstinos and the 
charged particles possess non-derivative interactions and as such it is 
compatible to identify the Goldstinos with the neutrinos.
This is 
in contrast to the situation considered in [\VA] 
where the electron and Goldstinos were both fundamental particles 
in the theory and their interaction can only involve 
derivative couplings. Thus when constructing theories of the standard model 
using wrapped branes one can identify the Goldstinos with the observed 
neutrinos and not be in contradiction with the low energy theorems.


\chapter{Conclusions}

In this paper we have analysed the interactions of Goldstone particles
with soliton states and showed that the scattering amplitudes need
not vanish in the limit of zero momentum. 
This allowed us to the resolve a conflict with
the duality between type IIA string theory and M-theory, in a sense
forming a test of this duality. This discussion also helps elucidate
the origin of the minimal coupling of soliton states 
to the worldvolume
gauge field.
We also discussed a phenomenological 
scenario in which neutrinos arise naturally as Goldstino particles for
broken supersymmetry but still have non-derivative couplings to the
charged fields. It seems reasonable to us that this situation will
occur rather generically in other brane world phenomenological
models.  

It is important to note that there are other problems that arise when the 
neutrino is identified with a Goldstone mode of broken supersymmetry
(e.g. see  [\Fayet]). For example since  the neutrino
carries lepton number then presumably 
so must the corresponding broken supersymmetry
generator. 
Therefore if the supersymmetry is spontaneously broken in the usual
sense one  expects that there will be 
massive states which  carry both baryon number 
and lepton number.  Such states  would then cause problems for  
phenomenology.
 However it is not clear that the
standard treatment of this issue  applies in the case of branes that
we have advocated here. For example if some of the supersymmetries
are non-linearly realised then there simply may not be any of the
corresponding  superpartners on the brane. Indeed this is the case for
D-branes where the broken sixteen supersymmetries are non-linearly
realised in the Dirac-Born-Infeld action and there are no
corresponding (massive)
superpartners on the worldvolume. Another possiblity is
that lepton number is an accidental symmetry, which perhaps arises as a
remenant of a discrete symmetry in the 
supersymmetric action of an unwrapped M-fivebrane.

Another problem that arises in the old  four-dimensional $N=1$ models 
where the
neutrino is identified as a Goldstino mode is that only one 
such particle, and not three, should appear. However, within the framework 
discussed here it is possible that several neutrinos 
are simply related to the breaking of additional 
supersymmetries.

In closing we note that there is now
strong experimental evidence that not all the
neutrinos are massless. However, their observed  masses are small
compared to the other scales in a brane world model and may be
neglected to a good approximation.
Nevertheless these non-zero masses must be
explained at least in principle and one might suppose this was due to
a small breaking of
supersymmetry in the bulk. Indeed such a small breaking in the bulk
spacetime
would lead to a non-zero value of the cosmological constant for which
there is also recent (although at this stage less reliable)
experimental evidence. It would
be interesting to see if these two mass scales could be related to
each other. Although these scales are still very different 
such a relation would presumably also involve the
compactification scale of the extra dimensions.
In other words one can ask the question as to
whether or not the small, but non-zero, 
cosmological constant is due to a small breaking of bulk supersymmetry
which in turn gives rise to a mass for any would-be Goldstino neutrinos?


\chapter{Acknowledgements}

We would like to thank A. Uranga for discussions 
and N. Manton for helpful communication on the Skyrme model.
We  would also like to thank the Theory Division at CERN for its
hospitality where this work was initiated. 
This work was supported in part by the two EU networks entitled  "On 
Integrability, Nonperturbative effects, and Symmetry in Quantum Field 
Theory"  (FMRX-CT96-0012) and "Superstrings" (HPRN-CT-2000-00122). 
It was also  supported by the PPARC special grant PPA/G/S/1998/0061
and N.D.L. is support by a PPARC fellowship.

\refout

\end